\documentclass[journal]{IEEEtranTIE}
\usepackage{graphicx}
\usepackage{cite}
\usepackage{picinpar}
\usepackage{amsmath}
\usepackage{url}
\usepackage{flushend}
\usepackage[latin1]{inputenc}
\usepackage{colortbl}
\usepackage{soul}
\usepackage{multirow}
\usepackage{pifont}
\usepackage{color}
\usepackage{alltt}
\usepackage[hidelinks]{hyperref}
\usepackage{enumerate}
\usepackage{siunitx}
\usepackage{url}
\usepackage{epstopdf}
\usepackage{pbox}
\usepackage{multirow}
\usepackage{upgreek}
\usepackage{textcomp,mathcomp}
\usepackage{booktabs}
\usepackage{graphicx}

\begin{document}
\title{Towards Large Scale Atomic Manufacturing: Heterodyne Grating Interferometer with Zero Dead-Zone}

\author{
	\vskip 1em	
	% First A. Author1, \emph{Student Membership},
	% Second B. Author2, \emph{Membership},
	% \\ and Third C. Author3, \emph{Membership}
Can Cui$^{\dagger}$, Lvye Gao$^{\dagger}$, Pengbo Zhao, Menghan Yang, Lifu Liu, Yu Ma, Guangyao Huang, Shengtong Wang, Linbin Luo and Xinghui Li*
	\thanks{
	
		Manuscript received October xx, 2024; revised Month xx, xxxx; accepted Month x, xxxx.
		This work was supported in part by National Natural Science Foundation of China under Grant 62275142, Shenzhen Stability Support Program Project under Grant WDZC20231124201906001 and Guangdong Basic and Applied Research Fund under Grant 2021B1515120007. (Can Cui and Lvye Gao are co-first authors.)(Corresponding author: Xinghui Li.)
		% (Authors' names and affiliation) First A. Author1 and Second B. Author2 are with the xxx Department, University of xxx, City, Zip code, Country, on leave from the National Institute for xxx, City, Zip code, Country (e-mail: author@domain.com). 
  
Can Cui, Lvye Gao, Pengbo Zhao, Menghan Yang, Lifu Liu, Yu Ma, Guangyao Huang, Shengtong Wang and Linbin Luo are with Shenzhen International Graduate School, Tsinghua University, Shenzhen, 518055, China. (Email: cuic23@mails.tsinghua.edu.cn).		
  
Xinghui Li is with Shenzhen International Graduate School, Tsinghua University, Shenzhen, 518055, China, and is also with Tsinghua Berkeley Shenzhen Institute, Tsinghua University, Shenzhen, 518055, China (Email: li.xinghui@sz.tsinghua.edu.cn).}}

\maketitle	
\begin{abstract}
This paper presents a novel heterodyne grating interferometer designed to meet the precise measurement requirements of next-generation lithography systems and large-scale atomic-level manufacturing. Utilizing a dual-frequency light source, the interferometer enables simultaneous measurement of three degrees of freedom. Key advancements include a compact zero Dead-Zone optical path configuration, significantly enhancing measurement reliability by mitigating the impact of light source fluctuations and air refractive index variations. A comprehensive crosstalk error analysis was conducted, resulting in a robust correction algorithm that reduces errors to below 5\%. Performance testing of the prototype, size of 90$\times$90$\times$40mm\(^3\), demonstrated exceptional resolution (0.25 nm in the XY-axis and 0.3 nm in the Z-axis), superior linearity (6.9e-5, 8.1e-5 and 16.2e-5 for the X, Y, and Z axes, respectively), high repeatability (0.8 nm/1000 nm for the three axes) and stability (20 nm for the XY-axis and 60 nm for the Z-axis over 1000 seconds). Comparative analysis with existing measurement sensors highlights the proposed method's significant advantages in integration, multidimensional capabilities, and is expected to be widely used in fields such as integrated circuits, atomic-level manufacturing and aerospace technology.
\end{abstract}


\markboth{IEEE TRANSACTIONS ON INDUSTRIAL ELECTRONICS}%
{}

\definecolor{limegreen}{rgb}{0.2, 0.8, 0.2}
\definecolor{forestgreen}{rgb}{0.13, 0.55, 0.13}
\definecolor{greenhtml}{rgb}{0.0, 0.5, 0.0}

\section{Introduction}

% \IEEEPARstart{W}{ith} the rapid development of the global semiconductor and electronic information industry, industrial production gradually reflects the development trend of intelligence, miniaturization and integration\cite{yang2021}. The parts of some precision industries such as photolithography machines and advanced machine tools have entered the atomic level, which also places extremely high requirements on industrial manufacturing\cite{jafari2021,gao2021}. Precision measurement technology, as a key guarantee in high-end industrial manufacturing, requires extremely high accuracy, which has reached sub-nanometer or even picometer levels in some fields\cite{schmidt2012,bai2023}. In addition, high-precision measurement technology also puts forward new requirements for environmental robustness, multi degree-of-freedom (DOF) expansion and error suppression\cite{dong2021,wang2024}.
\IEEEPARstart{W}{ith} the rapid development of the global semiconductor and electronic information industry, industrial production demonstrates a trend towards increased intelligence, miniaturization, and integration\cite{yang2021}. Components in precision industries, such as lithography machines, have achieved atomic-level precision, which imposes stringent demands on industrial manufacturing\cite{jafari2021,gao2021}. Precision measurement technology, a critical enabler of high-end industrial manufacturing, demands extremely high accuracy, reaching sub-nanometer or even picometer levels in certain fields\cite{schmidt2012}. In addition, high-precision measurement technology has introduced new demands for environmental robustness, multi degree-of-freedom (DOF) capability, and error elimination\cite{wang2024}.

Current precision measurement technologies include capacitive sensors \cite{ye2020}, time grating sensors \cite{wang2020}, laser interferometers \cite{sun2021}, and grating interferometers \cite{zhu2022}. Capacitive sensors achieve sub-nanometer accuracy but are limited to millimeter ranges, which restricts their use in larger-scale applications \cite{ma2023}. Time grating sensors offer linear displacement accuracy of about one hundred nanometers \cite{peng2023}. Laser interferometers provide sub-nanometer precision over meter distances but are affected by reflector quality and environmental disturbances, which leads to errors in multi-DOF configurations \cite{wang2022},\cite{zeng2015}. In contrast, grating interferometers offer greater robustness against environmental effects and can generate multi-level interference fringes, which makes them advantageous for high-precision, multi-DOF applications over extended measurement ranges \cite{kimura2012}.
Based on the light source type, grating interferometers can be divided into homodyne \cite{gao2007,lee2007,li2013} and heterodyne \cite{devine2009,yang2020} systems. Compared to homodyne systems, heterodyne systems offer higher accuracy and stronger anti-interference capability \cite{devine2009}. Therefore, heterodyne systems are particularly competitive in nanometer or sub-nanometer measurement scenarios. However, due to the extensive use of optical components, non-idealities introduce periodic nonlinear errors, affecting measurement accuracy \cite{yang2020,joo2020,hu2019}. There are two main sources of periodic nonlinear errors: frequency and polarization aliasing \cite{joo2020,hu2019}, and multi-level ghost reflections \cite{hu2015,fu2018}. Of these two errors, frequency and polarization aliasing has the greater impact.

\begin{figure*}[htbp]
\centering
\includegraphics[width=0.88\textwidth]{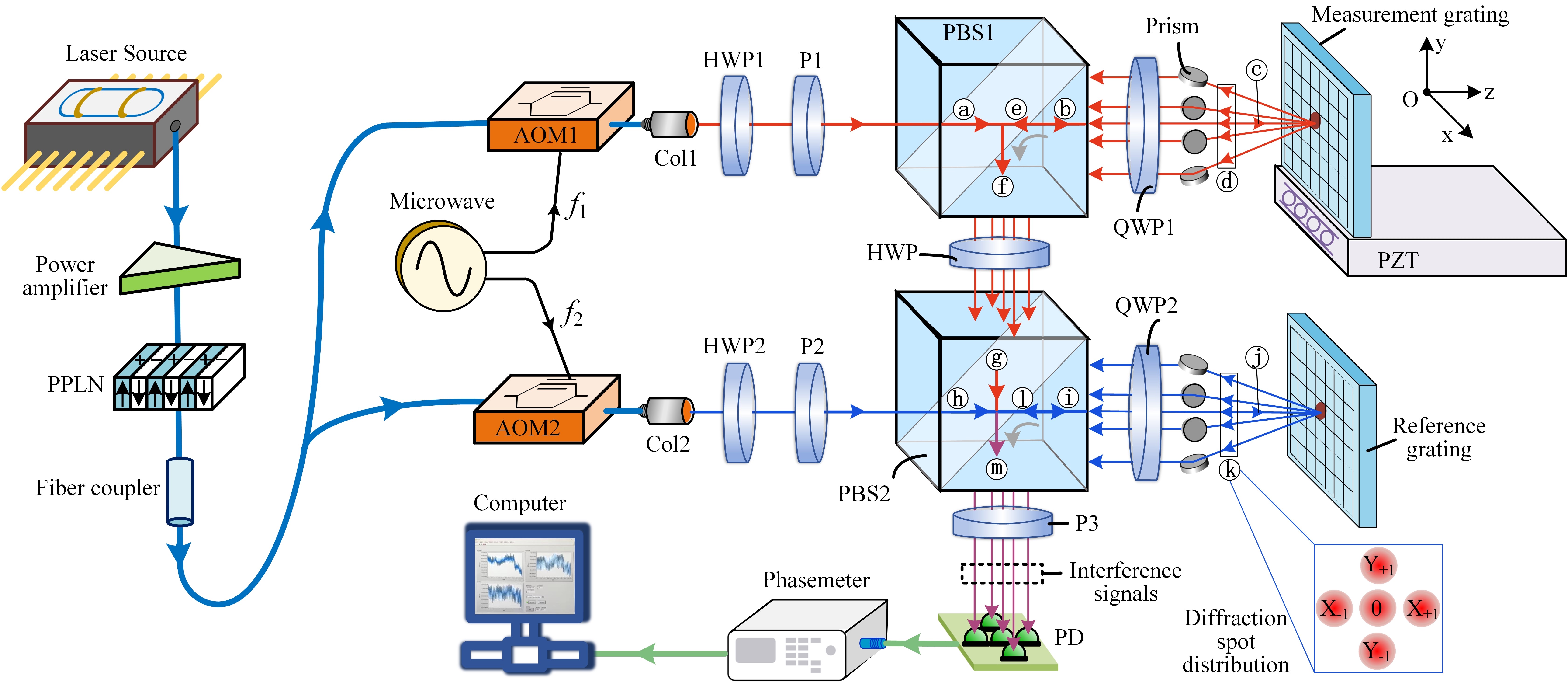}
\caption{Framework of grating interferometer system for 3-DOF measurement, where "a, b, c...m" represent the propagation order of the light path. PPLN: Periodically Poled Lithium Niobate; HWP: Half-Wave Plate; QWP: Quarter Wave Plate; COL: Fiber Coupler; AOM: Acousto-Optic Modulator; P: Polarizer; PBS: Polarizing Beam Splitter; PD: Photo Detector; PZT: Piezoelectric Stage.}
\label{fig_1}
\end{figure*}

To address the periodic nonlinear error caused by frequency aliasing of laser beams, extensive research has focused on two primary methods: spatially-separated structures and algorithm compensation. This paper focuses on the former, as spatially-separated structures can reduce errors at the source. Spatially-separated structures propagate dual-frequency light through distinct paths, effectively avoiding frequency and polarization aliasing encountered in traditional common-path systems and significantly minimizing periodic nonlinear errors. Wu et al. \cite{wu1999} first introduced a spatially separated heterodyne laser interferometer, reducing periodic nonlinear errors to 20 pm. This pioneering work inspired further research on spatially separated grating interferometers. For instance, Guan et al. \cite{guan2017} developed a differential interference heterodyne encoder that utilizes spatially separated input beams to mitigate periodic nonlinearity from polarization mixing. Subsequently, Xing et al. \cite{xing2017} proposed a new spatially separated heterodyne grating interferometer, further reducing periodic nonlinear errors by employing two modulated beams transmitted through separate optical fibers.

However, despite these advancements, spatially separated optical paths can induce a Dead-Zone effect due to differences in propagation distances. This effect may worsen measurement inaccuracies due to light source frequency instability and environmental noise. In response, some researchers have explored spatially separated structures designed to minimize Dead-Zone errors. Fu et al. \cite{fu2018-1} proposed a symmetric heterodyne interferometer that balances the probe and reference optical paths, effectively eliminating Dead-Zone effects and periodic nonlinear errors. More recently, Wang and Gao \cite{wang2024-1} developed a heterodyne grating interferometry system based on a quasi-common-path design. This innovative structure incorporates an oblique incidence design and wavelength stabilization, significantly reducing periodic nonlinear errors to 0.3 nm while enhancing resistance to interference and maintaining high resolution.
For in-plane and out-of-plane measurements, Hsieh et al. \cite{hsieh2013} designed a three-DOF grating interferometer utilizing a transmission two-dimensional grating, achieving measurement resolutions and ranges at the nanometer and millimeter levels, respectively. Lin et al. \cite{lin2017} introduced a wide-range three-axis grating encoder with nanometer resolution, enhancing the z-axis measurement range.

However, the aforementioned methods cannot simultaneously achieve both in-plane and out-of-plane measurements while maintaining low periodic nonlinear errors, and they struggle to meet the demands of future applications, such as large-scale atomic-level precision measurements.

In this article, driven by the need for large-scale atomic-level precision measurements, we present significant advancements in the design and functionality of a novel heterodyne grating interferometer. The main contributions of this work are summarized as follows:

\begin{enumerate} 
\item Development of a 3-DOF Heterodyne Grating Interferometer: We propose a heterodyne grating interferometer that employs a dual-frequency light source, enabling the simultaneous measurement of three degrees of freedom (DOF). A prototype has been successfully fabricated with dimensions of 90$\times$90$\times$40mm\(^3\), demonstrating the practicality and compactness of our design.

\item Zero Dead-Zone Optical Path Design: A zero Dead-Zone optical configuration has been meticulously designed to eliminate the Dead-Zone effect at its source, ensuring accurate displacement measurements. This enhancement significantly improves the system's resistance to disturbances caused by fluctuations in light source frequency and air refractive index, thereby ensuring more reliable measurement results.

\item Crosstalk Error Analysis and Robust Correction: We conducted a comprehensive analysis of crosstalk errors between different axes and developed a robust correction algorithm based on wavelet transform and Butterworth filtering. This approach successfully reduced crosstalk errors to below 5\%, significantly enhancing measurement accuracy.

\item Comprehensive Performance Testing of the Prototype: The integrated prototype underwent rigorous testing, evaluating key metrics including resolution, linearity, stability, and repeatability. Comparative analyses were performed against representative measurement devices from both industry and academia, demonstrating superior performance and reliability. 
\end{enumerate}
% These contributions collectively represent a substantial advancement in the field of optical grating interferometry, providing a foundation for future developments in high-precision measurement technologies.

% \begin{enumerate}[1)]
% 	\item Regular Papers and Special Section papers - Four to eight pages, including authors' bios and photos.
% 	\item Letters - One to three pages. Authors' bios and photos must not be included.
% \end{enumerate}

\section{Design and Measurement Principle}
% The schematic of the proposed quasi-common-path heterodyne grating interferometer is shown in Fig.. The system is consist of a dual-frequency laser system, a phase meter, a quasi-common-path optical structure and a Piezoactuator(PZT). The dual-frequency laser system emits two spatially separated laser beam with different frequency from two collimators(Col), respectively. The laser frequency shifts are caused by two acousto-optic modulators (AOM) modulated by the same micro-wave source and the output laser beams have no frequency aliasing. The following section will introduce the measurement principle of the grating interferometer and the quasi-common-path principle.
\subsection{Optical Design} 

The framework of the proposed 3-DOF zero Dead-Zone heterodyne grating interferometer is illustrated in Fig. \ref{fig_1}, and its components are described in detail below.

Light Source: The system utilizes a rubidium atomic seed laser of 1560 nm. This light is then amplified using a power amplifier and passed through Periodically Poled Lithium Niobate (PPLN) to convert the wavelength to 780 nm. Subsequently, the laser is coupled into an Acousto-Optic Modulator (AOM), generating two lasers with frequencies \( f_1 \) and \( f_2 \).

Reading Head: The two lasers, possessing a frequency difference \( f_0 \), are directed through Half-Wave Plate (HWP) and Polarizer (P) before entering Polarizing Beam Splitter (PBS). Each laser is then incident on both the measurement and the reference grating. The gratings diffract the incoming light into five beams: 0-th, \( X_{+1} \), \( X_{_{-1}} \), \( Y_{+1} \), and \( Y_{_{-1}} \). After passing through collimating modules, these beams return to the PBS. The specific propagation paths of the laser can be identified in Fig. \ref{fig_1}, denoted as ``a" to ``m". Ultimately, the ten diffracted beams interfere on a Photo Detector (PD) after traversing additional components, including HWP, QWP, and P, resulting in five interference signals.

Calculation Module: The five interference signals undergo signal conditioning and analog-to-digital conversion before being input into a phasemeter, which computes the 3-DOF displacement signals. The detailed calculation principles will be introduced in next section. The resulting 3-DOF displacement information, induced by the movement of the Piezoelectric Stage (PZT) and measurement grating, is displayed in real-time on the computer.

\subsection{Measurement Principle}
This section introduces how to solve the 3-DOF displacement of the measurement grating through the five interference signals $I_{0}$, $I_{X_{+1}}$, $I_{X_{-1}}$, $I_{Y_{+1}}$ and $I_{Y_{-1}}$ on the PD. Their expression are as follows:
\begin{equation}
	I_{i} \propto U_i \cos\left(2\pi{f_0}t + {\Phi}_{i}+ D_{i}\right)\\
 %        I_{X_{+1}} & \propto U_0 \cos\left[2\pi\left(f0\right)t + {\Phi}_{0} + D_{0}\right]\\
	% I_{X_{+1}} & \propto U_0 \cos\left[2\pi\left(f0\right)t + {\Phi}_{0}       + D_{0}\right]\\
 %        I_{X_{+1}} & \propto U_0 \cos\left[2\pi\left(f0\right)t + {\Phi}_{0} + D_{0}\right]\\	
 %        I_{X_{+1}} & \propto U_0 \cos\left[2\pi\left(f0\right)t + {\Phi}_{0} + D_{0}\right]\\
		% I_{PD2} & \propto U_0 \cos\left[2\pi\left(f_1-f_2\right)t + {\phi}_z+\phi_{k_2}\right]\\
		% I_{PD3} & \propto U_0 \cos\left[2\pi\left(f_1-f_2\right)t + {\Omega}_{x_{-1}} + {\Phi}_{x_{-1}} + \phi_{k_3}\right]\\
	\label{eq1}
\end{equation}
where $i$ represents $0$, $X_{+1}$, $X_{-1}$, $Y_{+1}$ and $Y_{-1}$, $I_{i}$ represents the intensity of the $i$-th interference signal, $U_i$ denotes the amplitude of the $i$-th signal. The variable $f_0$ is the frequency difference between the two laser beams. 
${\Phi}_{i}$ and $D_{i}$ represent the phase difference caused by the displacement of grating and the phase difference caused by the Dead-Zone, respectively.

According to the Doppler effect and the grating equation, the expression of ${\Phi}_{i}$ is as follows:
\begin{equation}
	\left\{	
	\begin{aligned}
		&\Phi_{0}=\frac{2\pi n}{\lambda}2z\\
		&\Phi_{X_{+1}}=\frac{2\pi n}{g}{x}+\frac{2\pi n}{\lambda}z\left(1+\cos{\theta}\right)\\
  		&\Phi_{X_{-1}}=\frac{-2\pi n}{g}{x}+\frac{2\pi n}{\lambda}z\left(1+\cos{\theta}\right)\\
    		&\Phi_{Y_{+1}}=\frac{2\pi n}{g}{y}+\frac{2\pi n}{\lambda}z\left(1+\cos{\theta}\right)\\
      		&\Phi_{Y_{-1}}=\frac{-2\pi n}{g}{y}+\frac{2\pi n}{\lambda}z\left(1+\cos{\theta}\right)\\
	\end{aligned}
	\right.
	\label{phase variation caused by displacement z1}
\end{equation}
where $\lambda$ represents the wavelength of the laser. $x$, $y$, and $z$ denote the displacements of the grating along the XYZ axis, respectively. $\theta$ is the 1-th diffraction angle of the grating.
$n$ is the refractive index of air. $g$ indicates the grating period. 

\begin{equation}
	\left\{	
	\begin{aligned}
&x = \frac{g}{4\pi n} \left( \Phi_{X_{+1}} - \Phi_{X_{-1}} \right)\\
&y = \frac{g}{4\pi n} \left( \Phi_{Y_{+1}} - \Phi_{Y_{-1}} \right)\\
&z = \frac{\lambda \left( \Phi_{X_{+1}} + \Phi_{X_{-1}} + \Phi_{Y_{+1}} + \Phi_{Y_{-1}} - 4\Phi_0 \right)}{8\pi n \left( 1 - \frac{\lambda^2}{g^2} \right)} \\
	\end{aligned}
	\right.
	\label{phase variation caused by displacement z}
\end{equation}
where the 1-th angle of the grating $\theta$ is represented by $\lambda$ and $g$ through the grating equation. At this point, the 3-DOF displacement of grating can be calculated from the phase change ${\Phi}_{i}$ in PD.

\subsection{Zero Dead-Zone Optical Path}

However, from Equation (\ref{eq1}), it is evident that the phase change at the PD consists of two components: the phase change \( \Phi \) caused by the displacement of the grating, and the additional phase \( D \) introduced by the Dead-Zone. 
To briefly explain the concept of the Dead-Zone: it arises due to the differing propagation distances of the two interfering signals in space, which leads to inconsistent effects from factors such as air refractive index variations and fluctuations in the laser frequency. As a result, an additional phase difference is created. Consequently, when calculating the phase of the signal at the PD, the presence of the Dead-Zone results in inaccuracies, ultimately causing displacement measurement errors.
The phase difference caused by the Dead-Zone can be expressed by the following formula:
\begin{equation}
	D_{i} = \frac{2\pi L_{i}  }{\lambda} \Delta n  + \frac{2 \pi n L_{i}}{c} {\Delta f}
	\label{Dead-Zone phase fluctuation}
\end{equation}
where $D_{i}$ represents the phase change of Dead-Zone. $L_{i}$ represents the length of Dead-Zone. $\Delta n$ represents the change in the refractive index of air, and $\Delta f$ represents the change in the frequency of the light source. Based on this, we quantified the measurement error caused by the Dead-Zone, as illustrated in the Fig. \ref{fig_deadzone} below:
\begin{figure}[htbp]
\centering
\includegraphics[width=0.7\columnwidth]{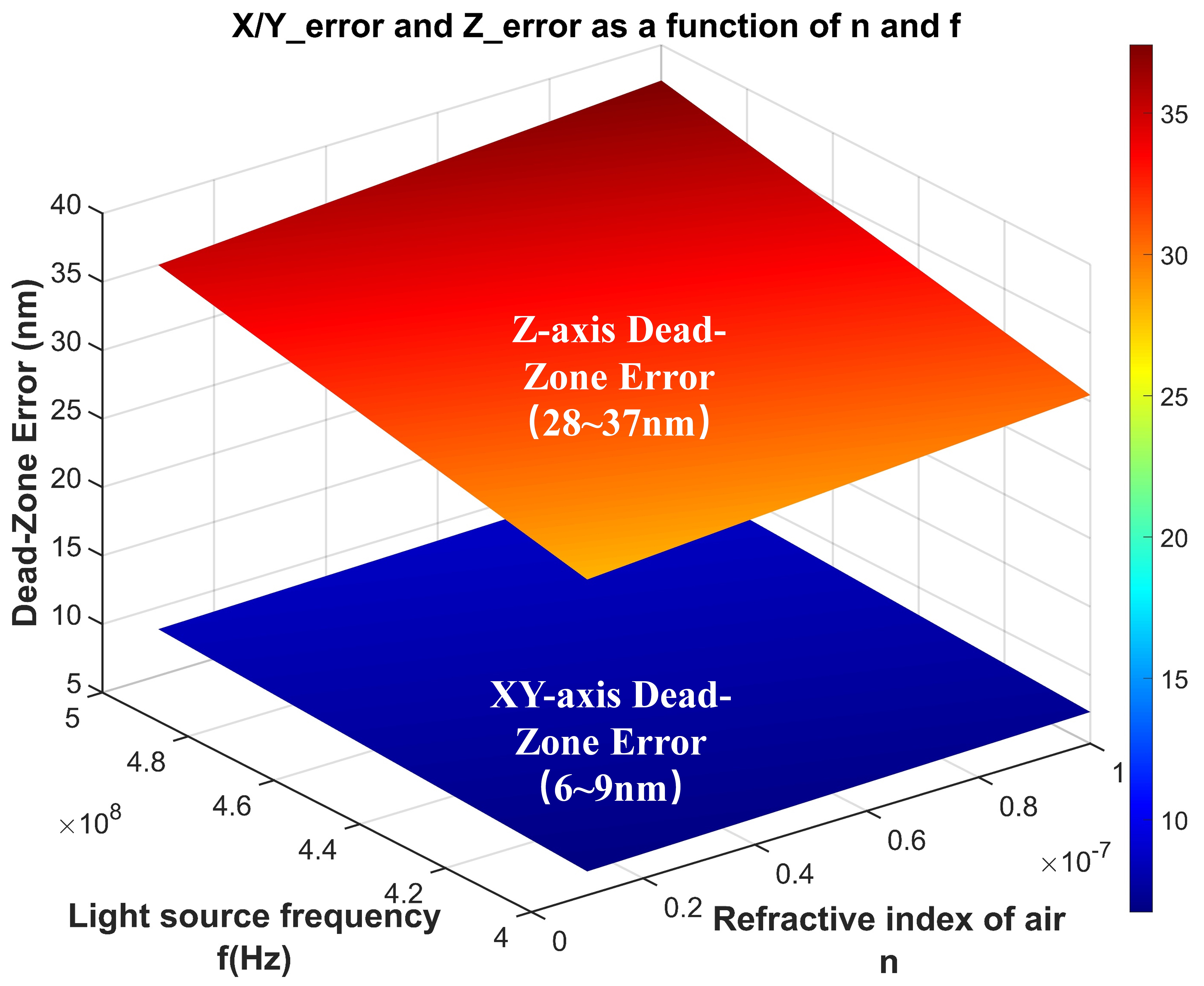}
\caption{XYZ-axis measurement error caused by Dead-Zone.}
\label{fig_deadzone}
\end{figure}
%在仿真中，死区的长度被设置为10mm，折射率的波动范围为(1e-8,1e-7),光源频率的波动范围为(400Mhz,500Mhz)，以上设置均来自实测数据。从图2可以看到，死区引起的XY的测量误差为6至9nm，而死区引起的Z的测量误差为28至37nm。总之，仅仅10mm的死区即可产生纳米级至几十nm级的误差，因此消除死区的影响对于高精度测量至关重要。

%基于此，本文精心设计了光路能够消除死区产生的影响，下面是详细论述。图3代表了xoz平面内的光路传播路径，基于此可以分别计算PDx+1，PD0，和PDx_{-1}（顺序为图3右侧PD从下至上）的光学死区，公式如下：

In the simulation, the length of the Dead-Zone is set to 10 mm, with fluctuations in the refractive index ranging from 1e-8 to 1e-7 and fluctuations in the laser frequency between 400 MHz and 500 MHz. These parameters are derived from experimental measurements. As shown in Fig. \ref{fig_deadzone}, the measurement error in the XY-axis  caused by the Dead-Zone ranges from 6 to 9 nm, while the measurement error in the Z-axis is between 28 and 37 nm. Overall, a mere 10 mm Dead-Zone can lead to measurement errors on the order of nanometers to several tens of nanometers. Therefore, mitigating the impact of the Dead-Zone is crucial for achieving high-precision measurements.

To address this issue, this study carefully designs the optical path to eliminate the effects of the Dead-Zone. Fig. \ref{fig_path} illustrates the propagation path of light within the XOZ plane, allowing for the calculation of the optical Dead-Zones for $PD_{X_{+1}}$, $PD_0$, and $PD_{X_{-1}}$ (from bottom to top on the right side of Fig. \ref{fig_path}). The formulas for these calculations are as follows:
\begin{figure}[htbp]
\centering
\includegraphics[width=0.9\columnwidth]{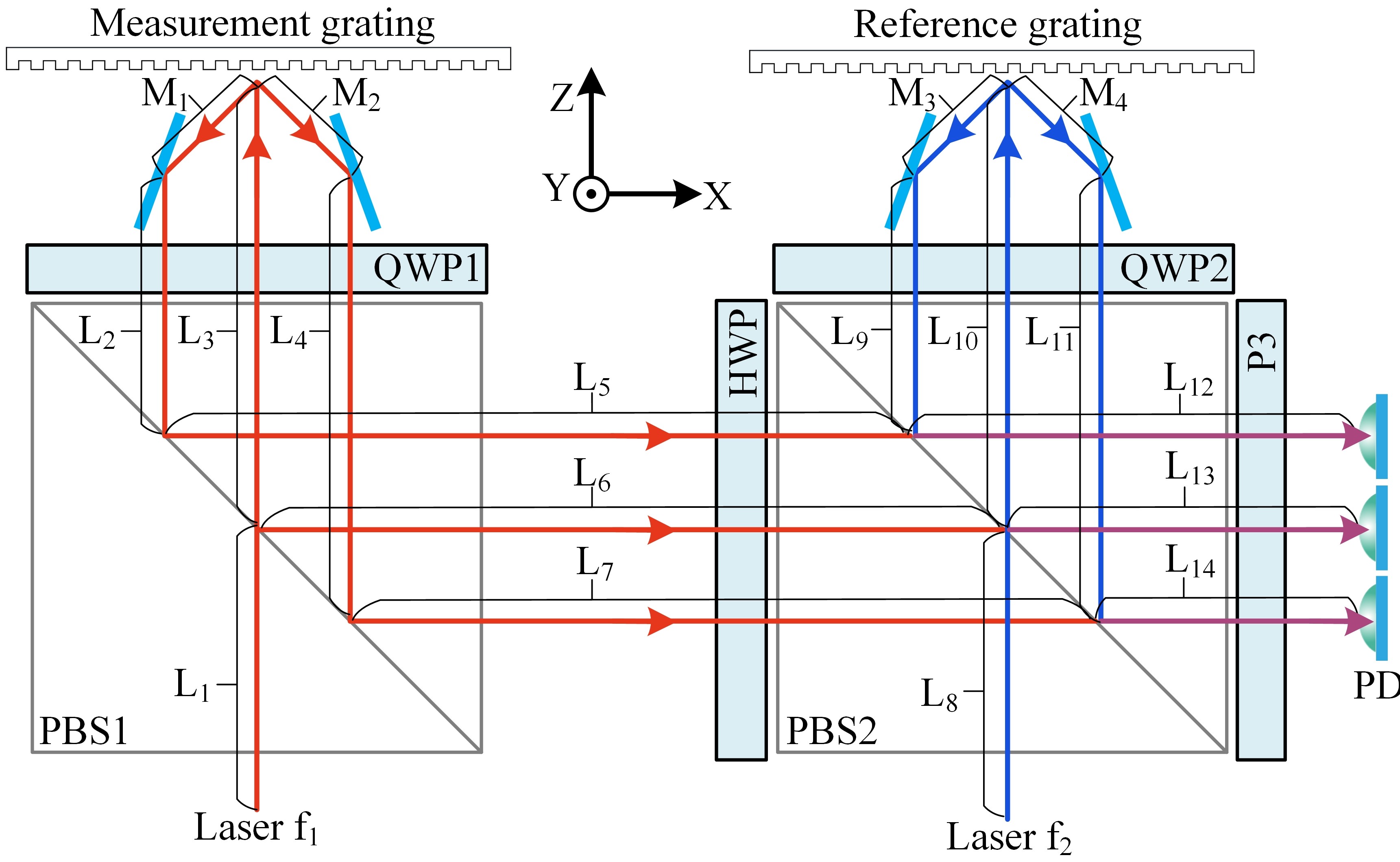}
\caption{Light propagation path in the XOZ plane.}
\label{fig_path}
\end{figure}

\begin{equation}
\left\{
\begin{aligned}
    % L_{x} &= (L_3+M_2+L_4+L_7+L_{14})-(L_3+L_4+L_9-L_3-L_6-L_{10}) \\
    % &\quad +(M_1+M_3-M_2-M_4) \\
    L_{X_{+1}} &= (L_1+L_3+M_2+L_4+L_7+L_{14})\\
    &\quad-(L_8+L_{10}+M_4+L_{11}+L_{14}) \\
    L_{X_{-1}} &= (L_1+L_3+M_1+L_2+L_5+L_{12})\\
    &\quad-(L_8+L_{10}+M_3+L_{9}+L_{12}) \\
    % L_{Y+1}, L_{Y_{-1}} &= L_{X+1}, L_{X_{-1}} \\
    L_{0} &= (L_1+L_3+L_3+L_6+L_{13})\\
    &\quad-(L_8+L_{10}+L_{10}+L_{13}) \\
    % L_{0} &=\frac{1}{2} \left[(M_1+M_2-M_3-M_4) \right.\\
    % &\quad +(L_2+L_3-L_9-L_{10}) \\
    % &\quad \left.+(L_4+L_6-2L_5)+(2L_8-2L_1)
    % \right]
\end{aligned}
\right.
\end{equation}
\begin{equation}
L_{Y_{+1}}, L_{Y_{-1}} = L_{X_{+1}}, L_{X_{-1}} \\
\end{equation}
% Dead-Zone $L_d$ is generally referred to the non-common optical path influenced by the environment and the so-called Dead-Zone phases caused by these can be denoted as $\phi_d$. Dead-Zone phases vary with the fluctuation of vacuum wavelength of laser and the refractive index of air. The influence of Dead-Zone can be denoted as:
\begin{equation}
	\left\{	
	\begin{aligned}
&x_{error} = \frac{g}{4\pi n} \left( D_{X_{+1}} - D_{X_{-1}} \right)\\
&y_{error} = \frac{g}{4\pi n} \left( D_{Y_{+1}} - D_{Y_{-1}} \right)\\
&z_{error} = \frac{\lambda \left( D_{X_{+1}} + D_{X_{-1}} + D_{Y_{+1}} + D_{Y_{-1}} - 4D_0 \right)}{8\pi n \left( 1 - \frac{\lambda^2}{g^2} \right)} \\
	\end{aligned}
	\right.
	\label{phase variation caused by displacement z33}
\end{equation}

where \( L \) and \( M \) are labeled in Fig.\ref{fig_path} and are arranged symmetrically. The variables \(x_{error}\), \(y_{error}\), and \(z_{error}\) represent the errors in the measurement due to the Dead-Zone. By simultaneously solving Equations (4), (5), (6), and (7), we can derive that \(x_{error}=y_{error}=z_{error}=0\). In short, the proposed optical path structure can solve the error caused by the Dead-Zone from the source, providing a basis for multi-DOF measurement at the sub-nanometer level or even the atomic-level.

\begin{figure*}[htbp]
\centering
\includegraphics[width=0.8\textwidth]{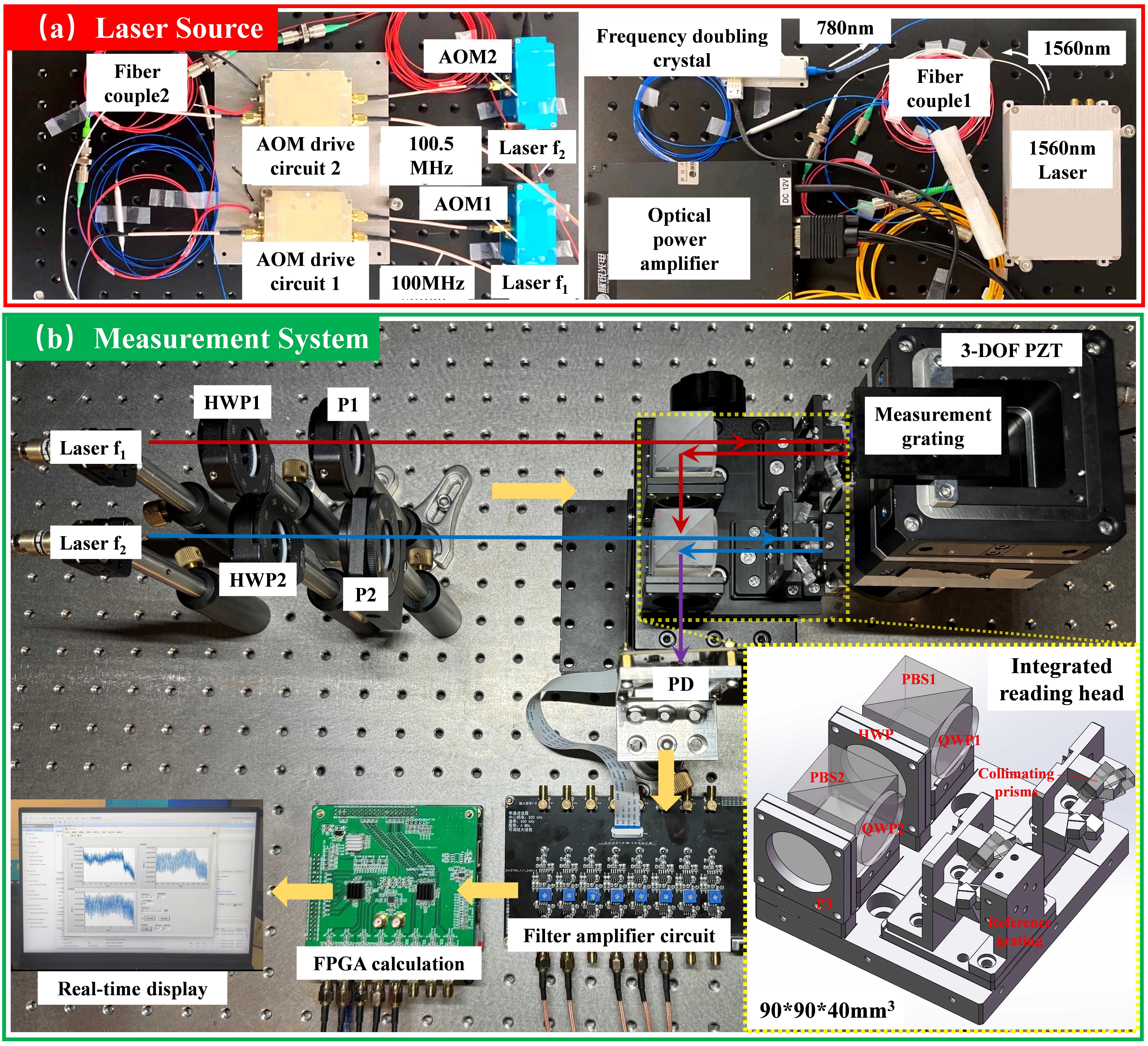}
\caption{Prototype of grating interferometer system for 3D measurement. (a) Heterodyne dual frequency light source. (b) Measurement system, including reading head, 2D grating, 3-DOF Piezoelectric Stage (PZT) and real-time processing display module.}
\label{fig_real}
\end{figure*}

\section{System and Experiments}
\subsection{Prototype System}

% This figure provides a detailed layout of a laser-based measurement system, showcasing the components involved in both the laser source generation and the measurement process. Key Sections of the System:
To verify the effectiveness and application ability of the proposed method, we built an experimental system, as shown in Fig. \ref{fig_real}. The following is a detailed description.

\subsubsection{Laser Source}
   As shown in Fig. \ref{fig_real}(a), the laser source section includes the generation and modulation of lasers. It features two main frequencies, \( f_1 \) and \( f_2 \), generated by the 1560 nm lasers. These lasers are frequency-doubled using a crystal to produce 780 nm light.
   The AOM1/AOM2 and their drive circuits (operating at 100 MHz and 100.5 MHz) are used to modulate the laser frequencies.
   Fiber coupling is employed to direct the laser beams from the source to the measurement system.
\begin{figure}[htbp]
\centering
\includegraphics[width=0.9\columnwidth]{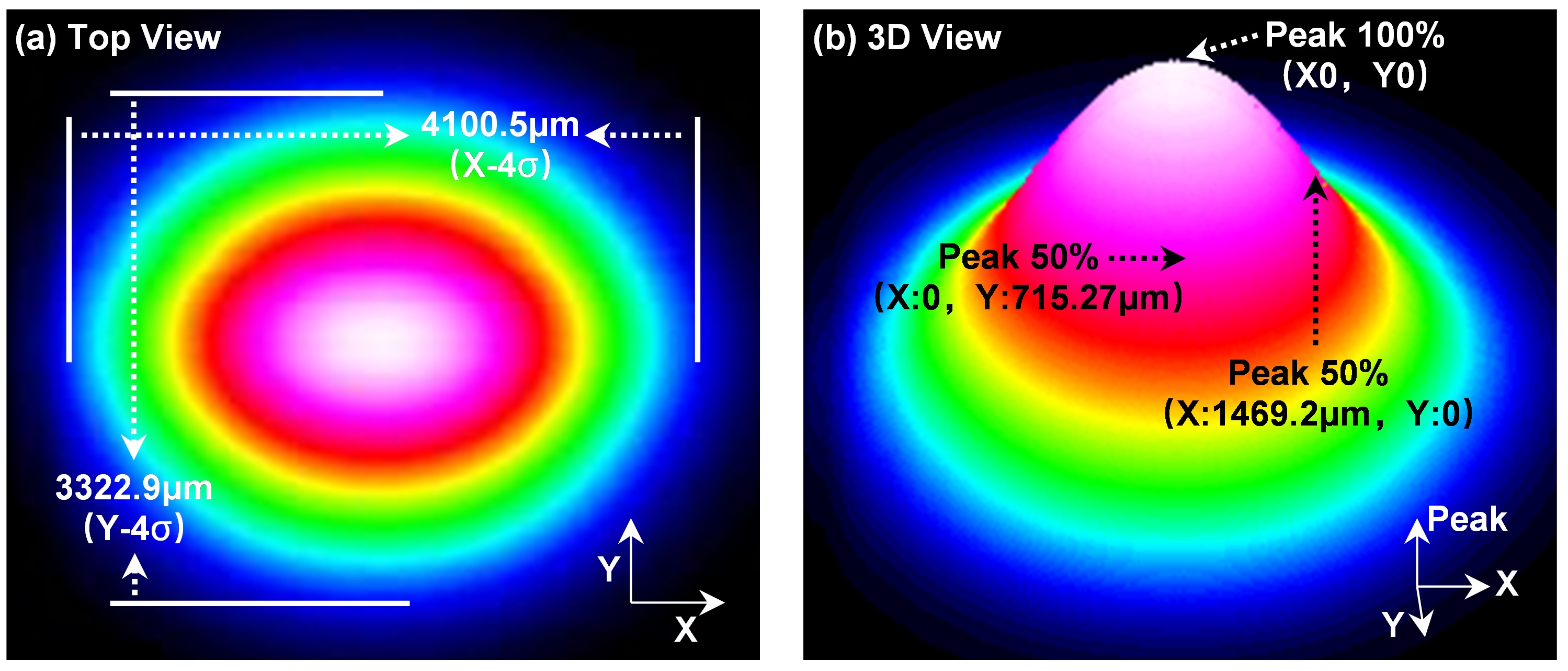}
\caption{Beam profiler test results of light sources. (a) Top view of Beam. (b) 3D view of Beam.}
\label{fig_light}
\end{figure}

In order to investigate the performance of the light source, we use a beam profiler to analyze the beam, as shown in Fig. \ref{fig_light}. As shown in Fig. \ref{fig_light}(a), the beam exhibits an elliptical intensity distribution, with widths of 4100.5$\upmu$m (X-4$\sigma$) and 3322.9$\upmu$m (Y-4$\sigma$).
The Fig. \ref{fig_light}(b) shows the maximum intensity peak is located at (0, 0) with a value of 100\%, the 50\% intensity points are at (0, 715.27$\upmu$m) and (1469.2$\upmu$m, 0).
In addition, the energy center stability of the beam is controlled at (3.36$\upmu$m,1.80$\upmu$m).
In short, the beam cross-section of the light source shows an elliptical Gaussian distribution, and has high stability, which is suitable for high precision measurement.

\subsubsection{Measurement System}
   The laser beams, \( f_1 \) and \( f_2 \), pass through a series of optical components such as HWP and P to control their polarization and energy.
   These beams then interact with a 3-DOF PZT that precisely positions the measurement grating.
   The reflected or diffracted beams are detected by a PD after passing through the integrated reading head, which contains components like PBS, QWP, and collimating prisms, all mounted on a compact 90$\times$90$\times$40mm$^{3}$ setup.
   The detected signals are amplified by the filter amplifier circuit and sent to an Field Programmable Gate Array (FPGA) for real-time calculation.
   The results of the measurements are displayed on a real-time display monitor, showing the 3-DOF displacement.
Several sets of experiments will be conducted to test the performance of the system, as detailed below.

\subsection{Performance Test and Results}

\subsubsection{Resolution}
\begin{figure}[htbp]
\centering
\includegraphics[width=0.49\columnwidth]{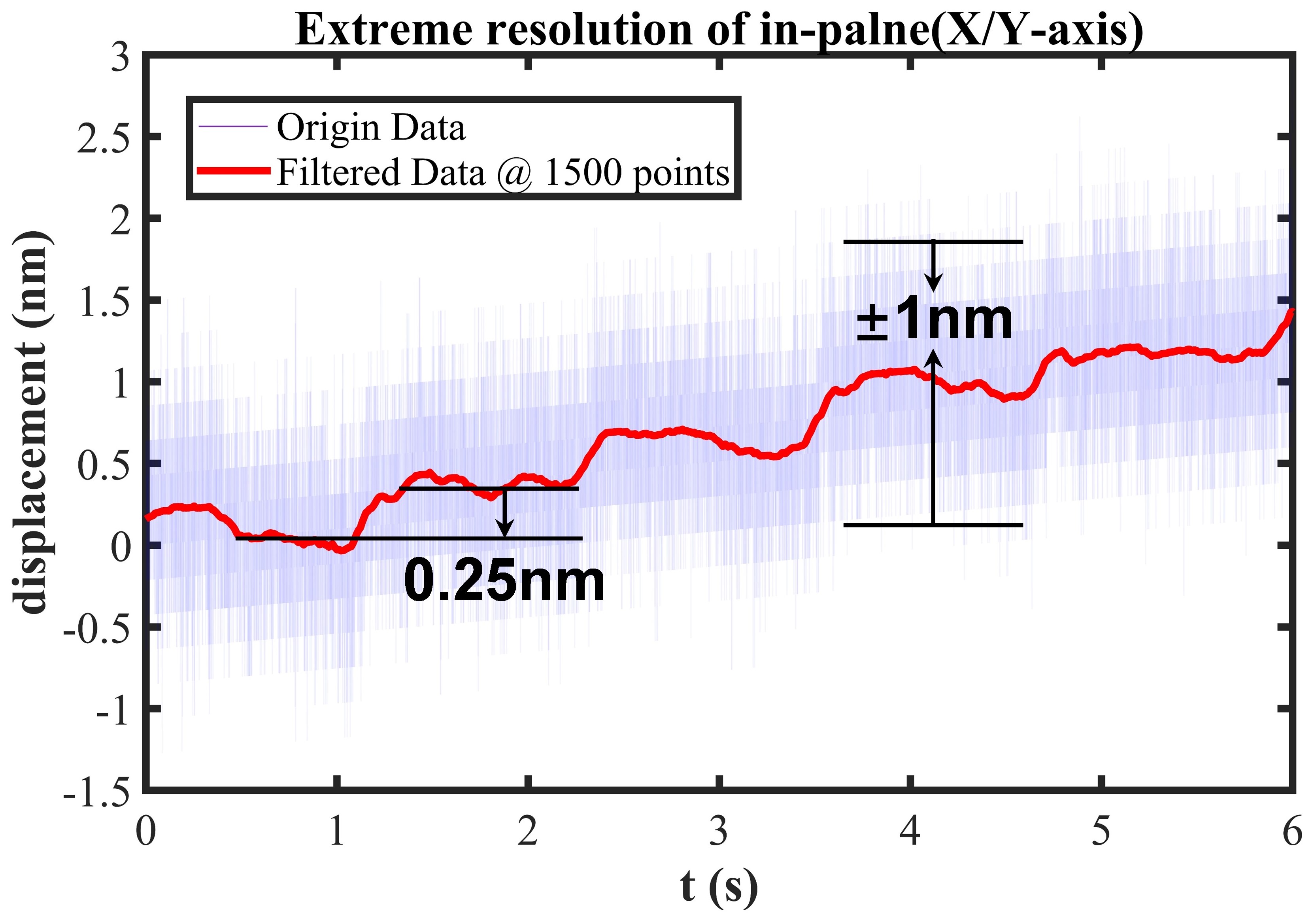}
\includegraphics[width=0.49\columnwidth]{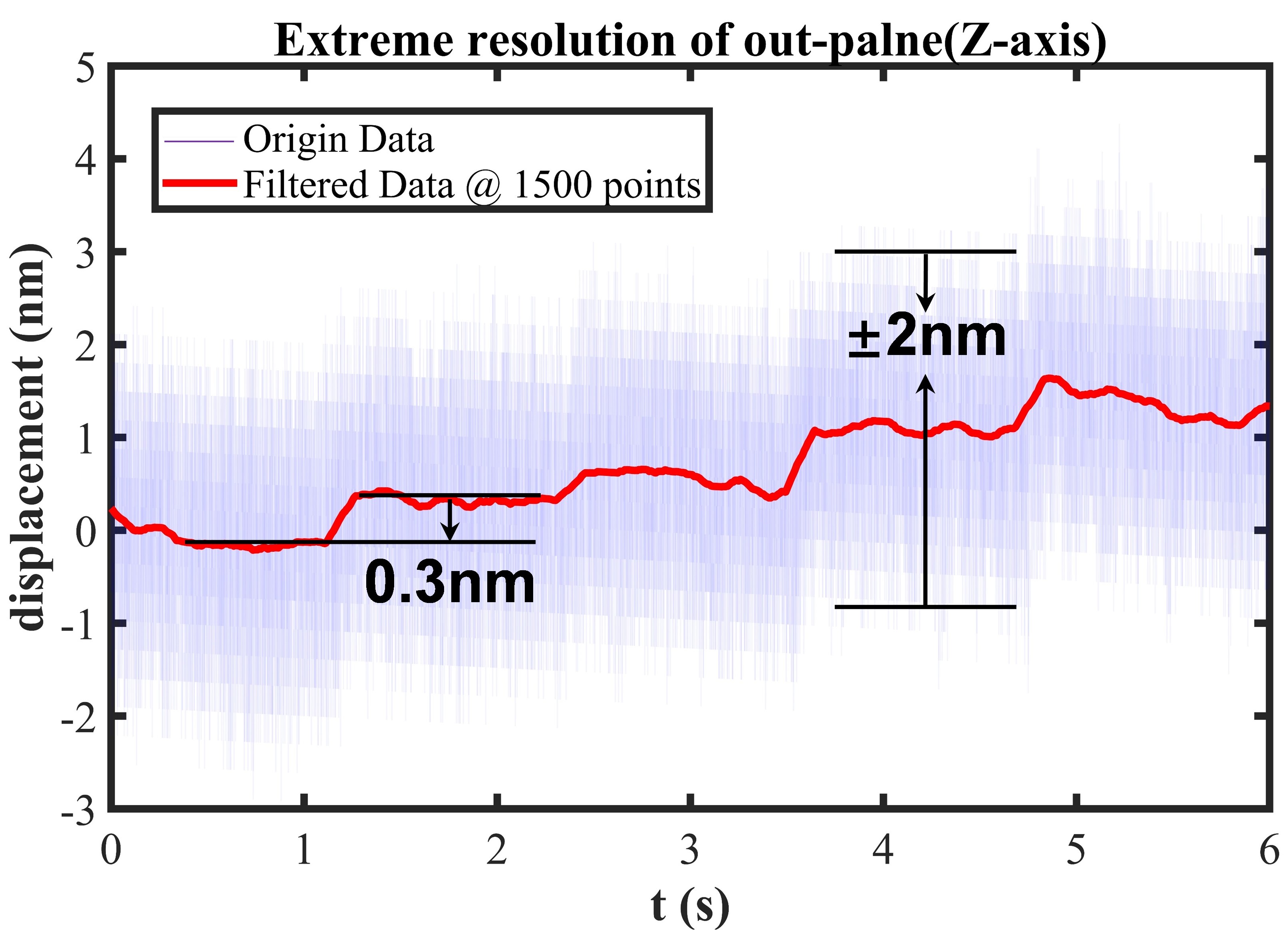}
\caption{Extreme resolution test. (Left) XY-axis. (Right) Z-axis.}
\label{fig_resolution1}
\end{figure}

\begin{figure}[htbp]
\centering
\includegraphics[width=0.95\columnwidth]{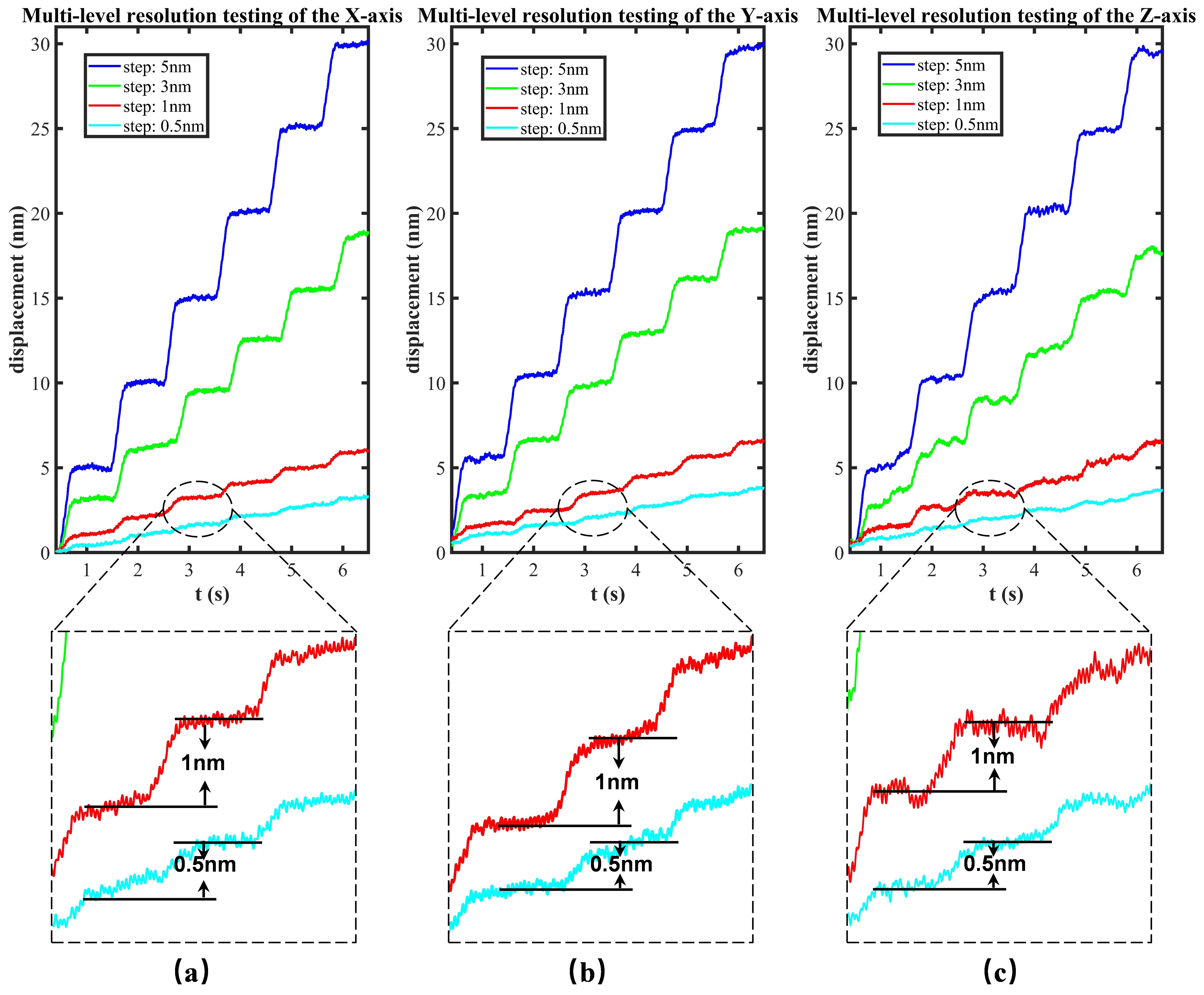}
\caption{Multi-level resolution test. (a) X-axis. (b) Y-axis. (c) Z-axis.}
\label{fig_resolution2}
\end{figure}

Resolution represents the ability of the system to distinguish the smallest displacement.
Fig. \ref{fig_resolution1} presents the extreme resolution tests for both in-plane (X/Y-axis) and out-plane (Z-axis) displacements. The left graph shows the in-plane measurement, with a resolution of 0.25 nm. The right graph depicts the out-plane measurement, with a resolution of 0.3 nm. Besides, Fig. \ref{fig_resolution2} displays multi-level resolution testing for the X, Y, and Z axis, containing step sizes of 0.5 nm, 1 nm, 3 nm, and 5 nm. 
In summary, the above experiments show that the system can resolve displacements at the near-atomic level (0.2 to 0.3nm), and has high universality for displacements of different sizes.

\subsubsection{Linearity}
\begin{figure}[htbp]
\centering
\includegraphics[width=0.9\columnwidth]{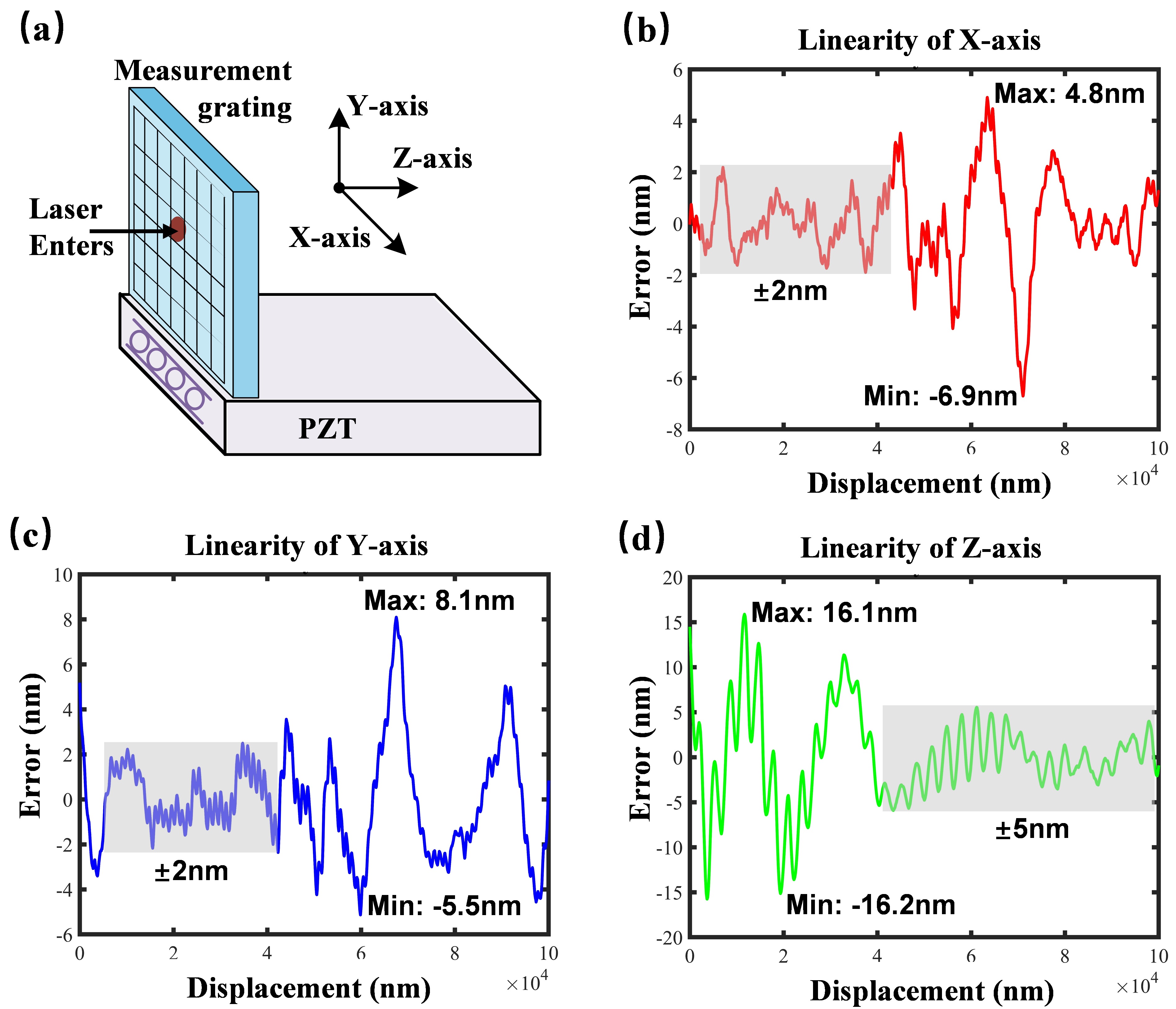}
\caption{Linearity test. (a) A diagram of the directions of different axis. (b) Linearity of X-axis. (c) Linearity of Y-axis. (d) Linearity of Z-axis.}
\label{fig_linear}
\end{figure}

Linearity represents the error level of the measuring system within a certain measuring range.
Fig. \ref{fig_linear}(a) represents the directions of different axis.
For the linearity of X-axis, as shown in Fig. \ref{fig_linear}(b), the maximum and minimum  error observed is 4.8nm and -6.9nm. The level of error in the stationary phase is 2nm. For the linearity of Y-axis, as shown in Fig. \ref{fig_linear}(c), the maximum and minimum  error observed is 8.1nm and -5.5nm. The level of error in the stationary phase is also 2nm. For the linearity of Z-axis, as shown in Fig. \ref{fig_linear}(d), the maximum and minimum  error observed is 16.1nm and -16.2nm. The level of error in the stationary phase is 5nm. 
So the linearity of the XYZ-axis is 6.9e-5/100$\upmu$m, 8.1e-5/100$\upmu$m and 16.2e-5/100$\upmu$m, respectively. In the stationary phase, the linearity of XYZ-axis can be increased to 2e-5/100$\upmu$m, 2e-5/100$\upmu$m and 5e-5/100$\upmu$m, respectively. The above experiments confirm the high precision of the system.

\begin{figure}[htbp]
\centering
\includegraphics[width=0.95\columnwidth]{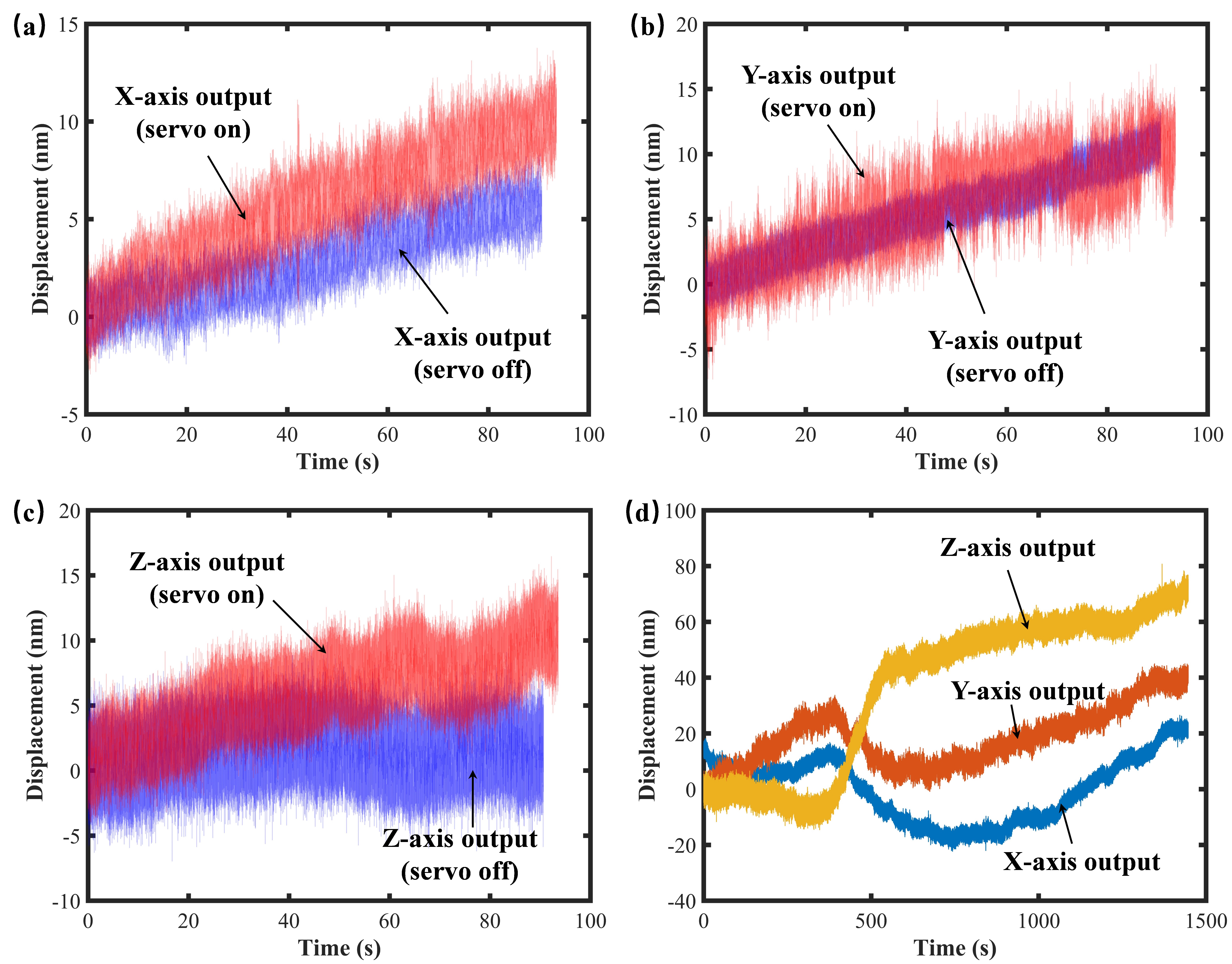}
\caption{Stability test. (a) X-axis open-loop and closed-loop 90s stability. (b) Y-axis open-loop and closed-loop 90s stability. (c) Z-axis open-loop and closed-loop 90s stability. (d) XYZ-axis closed-loop 1400s stability.}
\label{fig_sta}
\end{figure}

\subsubsection{Stability}
% \begin{figure*}[htbp]
% \centering
% \includegraphics[width=0.9\textwidth]{repeaty0910.jpg}
% \caption{Repeatability test. (a) X-axis. (b) Y-axis. (c) Z-axis.}
% \label{fig_repeat}
% \end{figure*}

Stability represents the ability of the system to operate stably, which is also related to environmental factors. Fig. \ref{fig_sta} shows the stability, where the servo on represents the open loop control, and the servo off represents the closed loop control.
The Fig. \ref{fig_sta}(a) shows the X-axis output shows distinct behavior under two conditions: with the servo on (red) and off (blue). The error steadily increases over time, with the servo on resulting in a maximum error of approximately 10nm, while the servo off exhibits a more erratic behavior with a fluctuation of about 5nm.
Similarly, as shown in Fig. \ref{fig_sta}(b), the Y-axis with servo-off provides a smooth displacement curve, reaching around 10 nm. 
As shown in Fig. \ref{fig_sta}(c), Z-axis output further reinforces these observations, with the servo on yielding a maximum error of 15 nm and a more consistent trend, whereas the servo off leads to maximum error of 5nm. Finally, as shown in Fig. \ref{fig_sta}(d), the outputs from all three axes are plotted over a longer time (1400s). The Z-axis (yellow) output exhibits significant fluctuations and reaches a displacement of around 80 nm, while the X-axis (blue) and Y-axis (red) outputs are more stable, highlighting the differences in stability across the axes under servo control. In general, the XYZ-axis of the system can maintain the stability of 20nm/1000s, 20nm/1000s and 60nm/1000s respectively at 1000s. 
\begin{figure}[htbp]
\centering
\includegraphics[width=0.9\columnwidth]{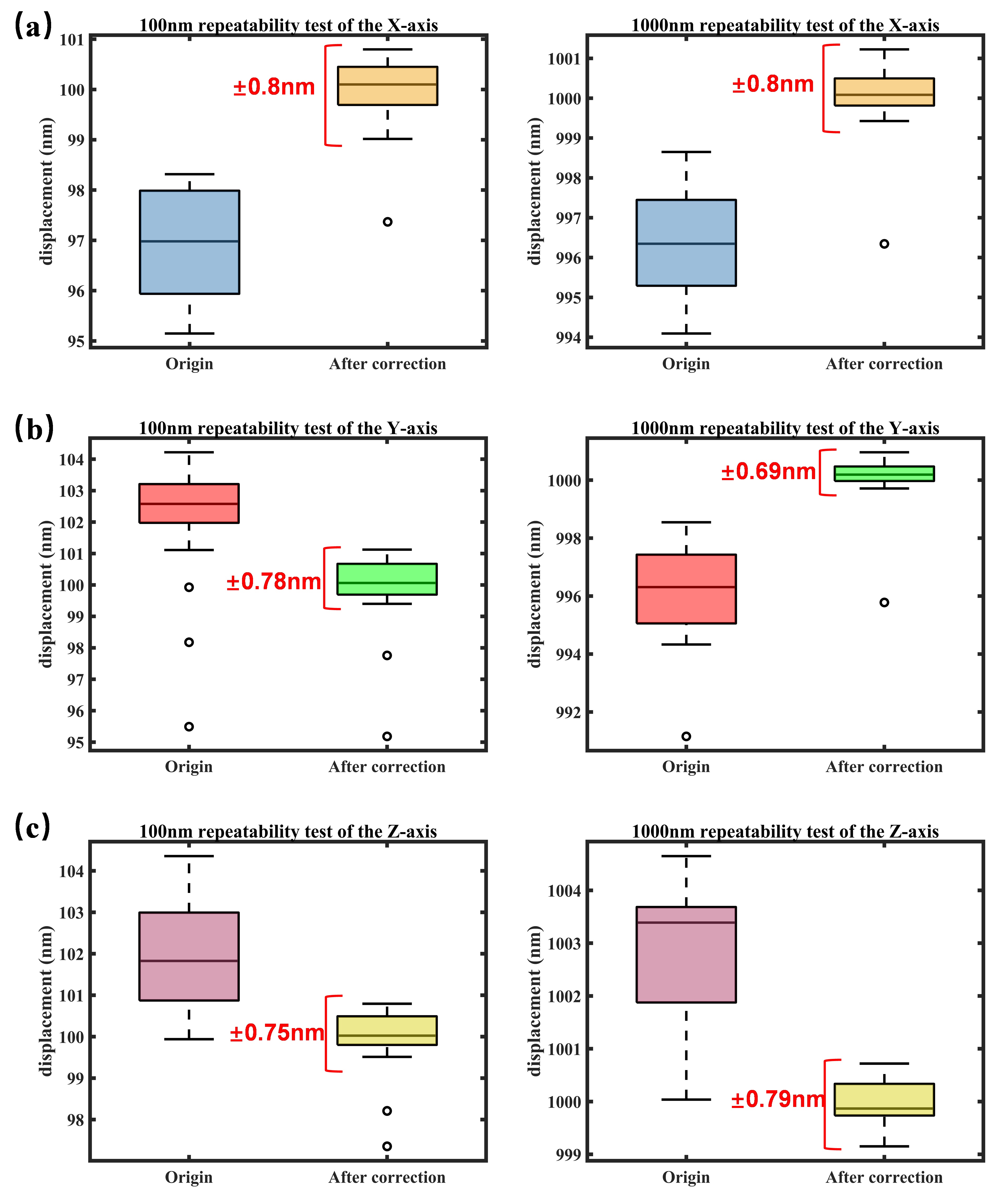}
\caption{Repeatability test. (a) X-axis. (b) Y-axis. (c) Z-axis.}
\label{fig_repeat}
\end{figure}
\subsubsection{Repeatability}
Repeatability refers to the ability of consistent with multiple measurements. As shown in Fig. \ref{fig_repeat}, the XYZ axis repeatability is characterized by boxplot, where correction stands for removing of cosine error. Fig. \ref{fig_repeat}(a) represents the X-axis repeatability. At 100nm and 1000nm, the repeatability is better than 0.8nm. The same conclusion can be drawn for the Y-axis and Z-axis, as shown in Fig. \ref{fig_repeat} (b) and (c). The above experiments prove that the system has excellent measurement repeatability, better than 0.8nm/1000nm.

\begin{figure*}[htbp]
\centering
\includegraphics[width=0.85\textwidth]{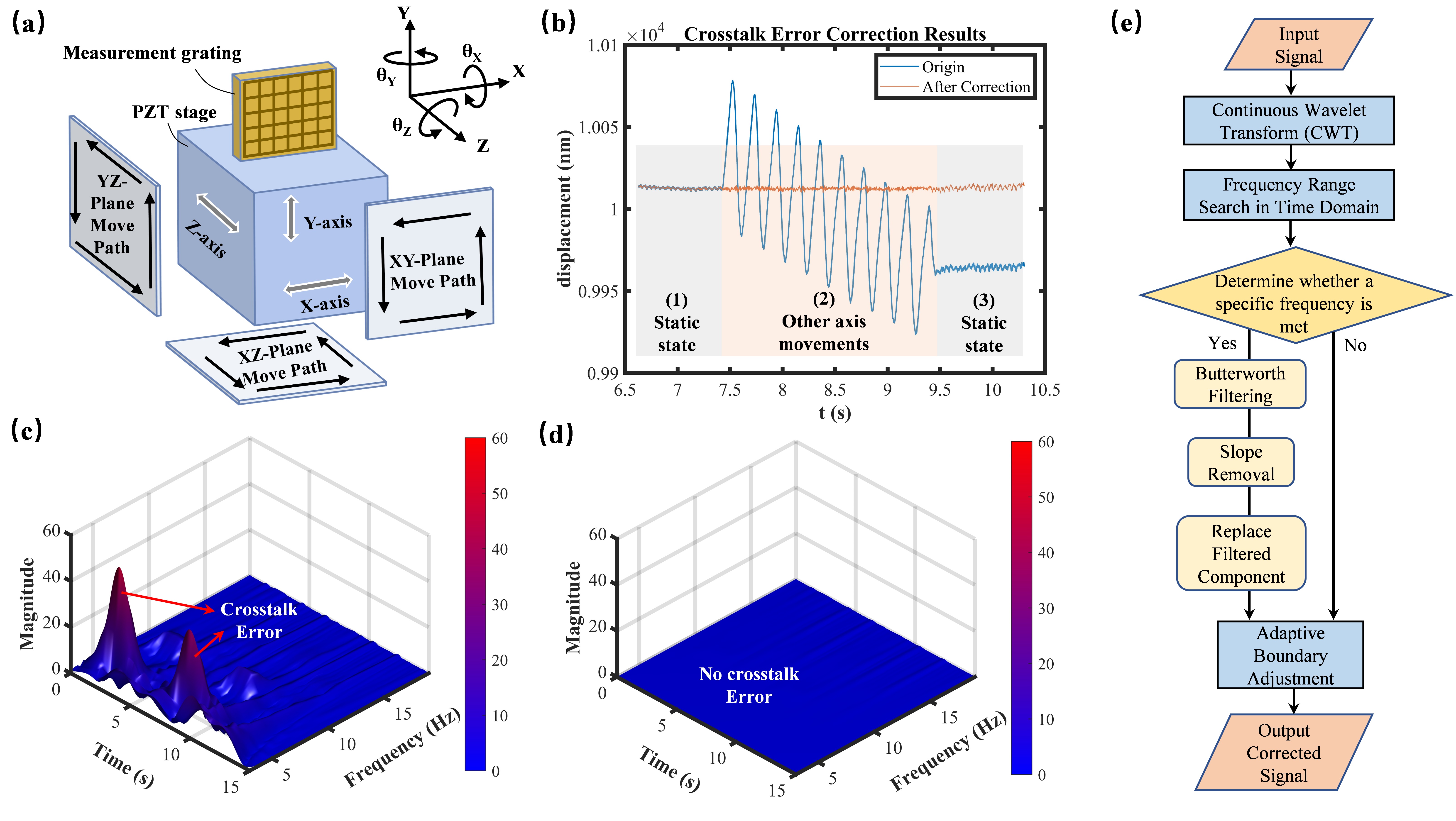}
\caption{Multi axes crosstalk error analysis and correction method. (a) The motion path of the grating in different planes. (b) In the XY plane, X-axis motion causes crosstalk error in the Y axis. (c) Wavelet transform diagram of crosstalk error. (d) Wavelet transform diagram of crosstalk error (after algorithm correction). (e) Flow chart of correction algorithm.}
\label{fig_cross}
\end{figure*}

\section{Error Analysis and Correction}
\subsection{Crosstalk Error Analysis}
For a measurement system, the crosstalk error between different axes is an important performance index, which is related to the ability of multi-axis synchronous measurement. 
Fig. \ref{fig_cross} represents the crosstalk error and corresponding correction methods.
Fig. \ref{fig_cross}(a) illustrates the measurement system configuration, highlighting how motion along the X-axis can induce crosstalk errors in the Y-axis. This phenomenon occurs primarily due to the interference of the laser as it traverses different axes, especially during simultaneous multi-axis movements.
There are many factors for this crosstalk phenomenon, including assembly errors between different axis, resonance phenomena, electronic noise and so on. Crosstalk error of X-axis motion on Y-axis can be shown in Fig. \ref{fig_cross}(b), The Y-axis output, while in X-axis static state, shows minimal error fluctuations. However, significant deviations occur during movements of X-axis, demonstrating the impact of crosstalk error. Post-correction data indicates a marked improvement in stability, confirming the effectiveness of the correction method.

\subsection{Correction Algorithm}
Fig. \ref{fig_cross} (c) and (d) provide the wavelet transform analysis of crosstalk errors before and after correction. The analysis reveals pronounced high-frequency components associated with crosstalk in Fig. \ref{fig_cross}(c), while Fig. \ref{fig_cross}(d) illustrates the absence of significant crosstalk errors post-correction. This indicates that the correction algorithm effectively identifies and mitigates crosstalk errors, enhancing measurement precision.
The correction algorithm flowchart in Fig. \ref{fig_cross}(e) outlines the systematic approach to address crosstalk errors. The process begins with the continuous wavelet transform (CWT) of the input signal, followed by a frequency range search in the time domain. Upon determining if a specific frequency criterion is met, the algorithm applies Butterworth filtering, removes any slope, and replaces filtered components. Finally, adaptive boundary adjustments are made to output the corrected signal. This structured methodology ensures effective control of crosstalk errors, significantly improving the reliability of the measurement system.
In conclusion, the implementation of correction algorithm enables the measurement system to achieve enhanced precision and stability.

\section{Conclusion}

% In this paper, we have developed a novel heterodyne grating interferometer tailored to meet the precise measurement demands of next-generation high-end lithography systems and large-scale atomic-level manufacturing. This innovative approach incorporates a zero Dead-Zone design that not only enhances performance but also achieves a compact integration, with size of 90$\times$90$\times$40mm\(^3\).
% The system demonstrates exceptional resolution, surpassing 0.25 nm in the XY-axis and 0.3 nm in the Z-axis. Moreover, linearity is maintained at superior levels, with performance metrics better than 6.9e-5, 8.1e-5 and 16.2e-5 for XYZ-axis, respectively. The repeatability of measurements has also been confirmed, with results exceeding 0.8nm/1000nm, 0.69nm/1000nm and 0.79nm/1000nm in the XYZ-axis. Notably, the system stability is better than 20nm in the XY-axis and better than 60nm in the Z-axis, at 1000 seconds test.

% Comparative analyses with representative measurement systems from both industry and academia, as detailed in Table I, reveal that the proposed method outperforms comprehensive competitiveness in terms of integration, multidimensional capabilities, resolution, linearity, stability, and repeatability. This advancement holds great promise for application in ultra-high precision measurement scenarios.

In this paper, we have developed a novel heterodyne grating interferometer tailored to meet the precise measurement demands of next-generation high-end lithography systems and large-scale atomic-level manufacturing. This innovative approach incorporates a zero Dead-Zone design that not only enhances performance but also achieves compact integration, with dimensions of 90$\times$90$\times$40mm\(^3\). The system demonstrates exceptional resolution, achieving 0.25 nm in the XY-axis and 0.3 nm in the Z-axis. Furthermore, linearity is maintained at superior levels, with performance metrics of 6.9e-5, 8.1e-5, and 16.2e-5 for the X, Y, and Z axis, respectively. The repeatability of measurements has also been confirmed, with results exceeding 0.8 nm/1000 nm, 0.69 nm/1000 nm, and 0.79 nm/1000 nm in the X, Y, and Z axis, respectively. Notably, the system's stability is better than 20 nm in the XY-axis and 60 nm in the Z-axis after a 1000-second test.

Comparative analyses with representative measurement systems from both industry and academia, reveal that the proposed method demonstrates superior competitiveness in terms of integration, multidimensional capabilities, resolution, linearity, stability, and repeatability. This advancement shows great potential for ultra-high precision measurement applications.

% References

\bibliographystyle{IEEEtranTIE}
\bibliography{main}\ 
% \end{IEEEbiography}
% \newpage

% %\vspace{-2cm}
% \begin{IEEEbiography}[{\includegraphics[width=1in,height=1.25in,clip,keepaspectratio]{photo-men.eps}}]
% {Third C. Author3} (M'99-SM'04-F'09) was born in City, Country. He received the M. and SM. and F. degrees in electrical engineering from University of City, Country in 1999, 2004 and 2009 respectively.

% The second paragraph uses the pronoun of the person (he or she) and not the author's last name. It lists military and work experience, including similar information to the previous author, including military, work experience, and other jobs.

% The third paragraph begins with the author's title and last name (e.g., Dr. Smith, Prof. Jones, Mr. Kajor, Ms. Hunter), including similar information to the previous author, including the list of any awards and work for IEEE committees and publications. The photograph is placed at the top left of the biography. Personal hobbies will be deleted from the biography.
% %\\ \\ 
% \end{IEEEbiography}

\end{document}